\documentclass[sigconf,nonacm]{acmart}
\AtBeginDocument{%
  }

\usepackage{dsfont}
\usepackage{tikz}
\usetikzlibrary{decorations.pathreplacing}

\renewcommand{\dblfloatpagefraction}{0.75}
\begin{document}

\title{Early Rug Pull Detection on Solana}

\author{
Saba Heydaridoost, Kasra Kashani, Mohammad Mowlavi Sorond, Behnam Bahrak\\
Tehran Institute for Advanced Studies (TeIAS), Khatam University, Tehran, Iran\\
\{s.heydaridoost, k.kashani, m.mowlavi, b.bahrak\}@teias.institute
}

\renewcommand{\shortauthors}{Heydaridoost et al.}

\begin{abstract}
 Web3 platforms through which users discover and trade newly launched crypto tokens, such as launchpads, decentralized-exchange (DEX) interfaces, wallets, and token-analytics sites, must assess each token while its legitimacy is still uncertain. On Solana, where launching a token costs little and tokens share standard programs rather than custom per-token smart contracts, rug pulls often unfold within minutes of creation, so a useful warning must rely on only a short history. Graph neural networks (GNNs) have improved rug-pull detection on EVM chains, but it remains unclear whether token-transfer structure adds predictive value over non-graph models at minute-scale decision points, and how this value changes as early history accumulates. Starting from 2,700 tokens sampled from SolRPDS, we collect one-minute market histories from Birdeye and token transfers from Solana RPC. Labels follow an explicit operational definition requiring a liquidity drain, a price collapse, and persistent liquidity depletion over a nominal 24-hour horizon after creation, while all features use only data observed at or before the prediction cutoff. We evaluate three observation windows of 12.5, 35.5, and 93 minutes, derived from the published lifespan distribution of Solana rug pulls and fixed before data collection, and represent each token by a temporal transfer graph over its window. Already within the first 12.5 minutes, rug-pull and non-rug transfer graphs differ in size, density, reciprocity, and receiver-side centralization. We compare GNNs with matched non-graph models that receive the same token-level and graph-summary features on identical samples, including a fixed panel of tokens observed in all three windows that separates the effect of longer observation from changes in the token population. The results provide practical guidance for designing early-warning mechanisms in Web3 interfaces for newly launched Solana tokens.
\end{abstract}



\maketitle

\section{Introduction}
\label{sec:intro}

Solana has become one of the most active ecosystems for launching and trading new
crypto tokens~\citep{chen2026hypecollapse}. A transaction costs a fraction of a cent
(about \$0.00025 on average~\citep{alhaidari2025}), every token is issued
through the same shared token program without a contract of its own, and launchpads
such as pump.fun\footnote{\url{https://pump.fun/}} let anyone create a tradable token
in a few steps. The same openness makes Solana a prime venue for \emph{rug pulls},
scams in which the creators of a token attract buyers and then withdraw its
liquidity or sell off their holdings, leaving investors with tokens they can no
longer sell. SolRPDS, built from 3.69~billion Solana transactions between
February 2021 and November 2024, flags 62{,}895 liquidity pools
as suspicious and finds rug-pull patterns in 22{,}195
tokens~\citep{alhaidari2025}, and \citet{chen2026hypecollapse} identify
76{,}469 rug-pull tokens in the first half of 2025 alone. These scams
are also fast: the median time from token creation to the first fraud-related
profit-taking is 35.5~minutes~\citep{chen2026hypecollapse}, so a warning
that reaches users in time must be based on the first minutes of a token's life.

Early rug-pull prediction from the first minutes of trading is feasible with
tabular features and tree ensembles~\citep{kalacheva2025memecoins,
yaremus2025detecting,li2026catching}, and on EVM chains, graph neural networks
(GNNs) over transaction graphs outperform feature-based
detectors~\citep{wu2024tokenscout,li2025hnn4rp,wu2025rphunter}. Whether the
transfer graph also helps at minute-scale decision points on Solana is open:
\citet{li2026catching} name temporal GNNs as future work, and reported graph gains
are usually measured against baselines with different features, populations, or
prediction points, so they cannot be attributed to relational structure alone. For
a platform, the practical question is about time: \emph{does the transfer graph let
a rug pull be detected earlier than a model without it?}

We answer this question with a controlled design. Starting from 2{,}700
tokens sampled from SolRPDS~\citep{alhaidari2025}, we collect one-minute market
histories and token transfers, label each token with an explicit operational
definition evaluated after the prediction cutoff, and observe it for 12.5,
35.5, and 93~minutes, the quartiles of the Solana rug-pull lifespan
distribution of \citet{chen2026hypecollapse}, fixed before data collection. GNNs and
non-graph models receive the same fourteen features and are compared on
identical test tokens, both in the full population of each window and in a fixed
panel of tokens observed in all windows. We address two research questions:
\begin{itemize}
  \item[\textbf{RQ1}] \emph{Across windows:} does a GNN at $q_1$ outperform
  XGBoost at $q_1$, $q_2$, or $q_3$, i.e., detect rug pulls earlier? We
  pre-specified this comparison as our primary analysis before evaluating on the
  test data.
  \item[\textbf{RQ2}] \emph{Within a window:} does a GNN outperform a non-graph model
  that has observed the token for the same time?
\end{itemize}

Our results show that the early signal is real, but we find no evidence that the
transfer graph makes detection earlier. Rug-pull and non-rug transfer graphs
already differ clearly after 12.5~minutes, most strongly in receiver-side
centralization. However, none of the three GNNs at $q_1$ is significantly better
than XGBoost at any of the three cutoffs (RQ1), and XGBoost itself does not improve
with longer observation: on the fixed panel, its test AP decreases from
0.464 at $q_1$ to 0.393 at $q_3$. Within each window, the selected GNN
reaches a higher AP than the selected non-graph model in 5 of 6
settings at a similar ROC-AUC (RQ2), but no difference remains significant after
correction. A power analysis shows that our test sets can only detect AP
differences of about 0.10--0.18, so these results are inconclusive
rather than evidence of no effect. Direction-aware, bagged, stacked, global-graph,
and heterogeneous-graph variants do not change this picture.

Our contributions are:
\begin{itemize}
  \item A leakage-aware Solana benchmark and protocol for early rug-pull detection:
  cutoff-separated operational labels, windows fixed in advance, a chronological
  split with a label-availability embargo, and full and panel evaluation scopes.
  \item Evidence that rug-pull and non-rug transfer graphs differ structurally
  within the first 12.5~minutes.
  \item A pre-specified comparison of graph and non-graph models with identical
    features and tokens across observation windows, with ablations.
  \item Practical implications for early-warning mechanisms on Web3 platforms.
\end{itemize}

\section{Related Work}
\label{sec:related}

\paragraph{Early rug-pull detection.}
\citet{mazorra2022donot} predicted rug pulls on Uniswap with a separate model for
each of the first 24 hours after pool creation and found that useful signals
emerge early and improve with longer histories. Later work used much shorter
windows: the first five minutes of trading suffice for useful predictions on
Ethereum~\citep{kalacheva2025memecoins} and TON~\citep{yaremus2025detecting}. On Solana, \citet{chen2026hypecollapse} characterize 76{,}469 rug-pull tokens
from the first half of 2025, including their lifecycles and mechanisms such as
freeze-authority abuse, but do not evaluate early prediction. \citet{li2026catching} predict
rug pulls on PumpFun and Raydium from the first five minutes, find tree ensembles
to be strong baselines, and name temporal GNNs as future work.
MemeTrans~\citep{hu2026memetrans} evaluates tabular and time-series models on
engineered launch features, and \citet{li2025trust} describe structural differences
between the transaction networks of rug-pull and sustainable meme coins without
evaluating graph-based prediction. SolRPDS~\citep{alhaidari2025} provides the token
identifiers from which our collection starts.

\paragraph{Graph-based detection.}
Graph-based methods have also been studied for early scam-token detection.
TokenScout~\citep{wu2024tokenscout} applies temporal graph learning to ERC-20
transfer multigraphs to identify scam tokens before scam execution, but leaves
evaluation at different lifecycle stages to future work.
HNN4RP~\citep{li2025hnn4rp} combines a heterogeneous transaction graph with
temporal fusion, RUGSCREENER~\citep{wu2025rugscreener} applies a temporal GNN, and
RPHunter~\citep{wu2025rphunter} fuses transaction graphs with smart-contract code
graphs and varies the number of observed transactions rather than the elapsed time.
Several of these methods draw on contract code, which has no per-token counterpart
on Solana (Section~\ref{subsec:solana}). More importantly, their graph and non-graph
models differ in features, populations, or prediction points, so the gain cannot be
attributed to relational structure, and none evaluates fixed minute-scale windows.
In the broader Web literature, PC-GNN~\citep{liu2021pcgnn} addresses imbalanced
graph fraud detection, Bi-GCN~\citep{bian2020bigcn} evaluates rumor detection at
early deadlines using only propagation observed before each deadline, and
\citet{huynh2025serial} reconstruct coordinated rug-pull operations from
cross-address transaction flows.

\paragraph{Temporal validity.}
\citet{shoaei2026tmrugpull} identify temporal leakage as a major weakness of
rug-pull datasets. We follow their principle by separating the information
available at the cutoff from the evidence that defines the outcome. To the best of our
knowledge, no prior work compares graph and non-graph models with shared token-level
and graph-summary predictors on the same Solana tokens across fixed early windows,
while GNNs additionally use endpoint features and connectivity.

\section{Background and Problem Formulation}
\label{sec:background}

\subsection{Solana Token and Account Model}
\label{subsec:solana}

On Ethereum, each token is a contract with its own code and
storage~\citep{ethereumAccounts2026}. Solana instead separates executable programs
from the accounts that hold state~\citep{solanaCore2026}: tokens are managed by the
shared Token or Token-2022 program, each token is identified by a mint account that
stores its supply, decimals, and optional mint and freeze authorities, and balances
are held in token accounts, each with an owner~\citep{solanaTokens2026}. Methods
that analyze per-token contract code~\citep{lin2024crpwarner,wu2025rphunter}
therefore do not transfer directly. Graph nodes represent token-account owners when available and token accounts
otherwise; we refer to both as \emph{transfer endpoints}.

\subsection{Temporal Transfer Graph}
\label{subsec:prelim-graph}

For token $i$ and window $q$, the recovered transfers form a directed temporal
multigraph $G_{i,q} = (V_{i,q}, E_{i,q}, T_{i,q}, X^V_{i,q}, X^E_{i,q})$, where
$V_{i,q}$ are transfer endpoints, $E_{i,q}$ are retained transfers, $T_{i,q}$ their
timestamps, and $X^V_{i,q}$ and $X^E_{i,q}$ node and edge attributes, as in earlier
temporal transfer-graph work~\citep{wu2024tokenscout}. Parallel edges between the
same ordered pair are kept, and the mint identifies the graph but is not a node.
Because transfers are discovered through the mint's transaction history, we treat
$G_{i,q}$ as a \emph{partial mint-visible transfer graph}.

\subsection{Problem Formulation}
\label{subsec:problem}

\begin{figure}[t]
  \centering
  \begin{tikzpicture}[x=0.72cm, >=stealth, font=\footnotesize]
    \fill[blue!12] (0,-0.22) rectangle (2.5,0.22);
    \draw[->, thick] (0,0) -- (10.6,0) node[right] {$t$};
    \foreach \x/\lab in {0/{t_0}, 2.5/{t_q}, 5.5/{\tau}, 7/{\tau+\Delta}, 10/{t_0+H}}
      \draw (\x,0.12) -- (\x,-0.12) node[below] {$\lab$};
    \fill[red!70!black] (5.5,0) circle (2pt);
    \draw[decorate, decoration={brace, amplitude=4pt}]
      (0,0.35) -- (2.5,0.35)
      node[midway, above=5pt, align=center] {predictors};
    \draw[decorate, decoration={brace, amplitude=4pt}]
      (2.55,0.35) -- (10,0.35)
      node[midway, above=5pt, align=center] {label evidence};
  \end{tikzpicture}
  \caption{Early-prediction setting for one token and window $q$. Predictors use
  only data up to the cutoff $t_q$. The rug-pull event at $\tau$ (the liquidity
  drop) and its confirmation (Section~\ref{subsec:labeling}) lie after the cutoff
  and within the outcome horizon $H$.}
  \Description{A horizontal time axis from token creation to 24 hours later. The
  interval from creation to the cutoff is shaded and marked as the source of
  predictors. The interval after the cutoff contains the rug-pull event and its
  persistence check, marked as label evidence.}
  \label{fig:timeline}
\end{figure}
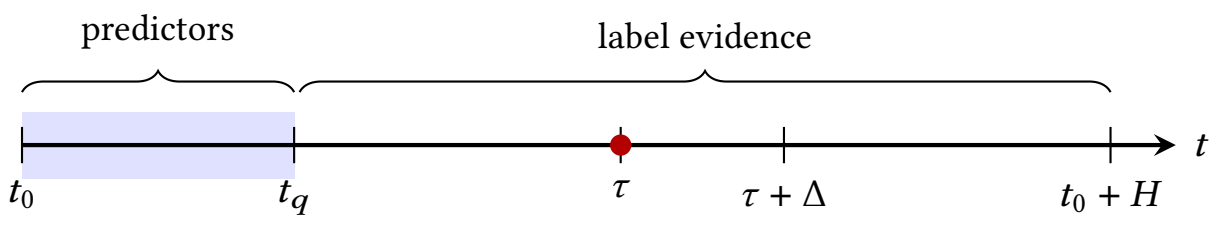

Let $t_0$ be the creation time of token $i$ and $w_q$ the length of window $q$,
with cutoff $t_q = t_0 + w_q$. The window length $w_q$ sets how early a prediction
is made; we compare model performance at fixed cutoffs rather than measuring when a
model first raises an alert. At $t_q$, a model must predict whether the token will
be a rug pull using token-level market features and the transfer graph $G_{i,q}$
from $[t_0, t_q)$, except that cutoff liquidity may be observed exactly at $t_q$.
The outcome is evaluated over a fixed nominal horizon $H = 24$~hours after
creation, identical for all windows, and $y_{i,q} \in \{0,1\}$ indicates whether
the token satisfies our operational rug-pull definition
(Section~\ref{subsec:labeling}) with the rug-pull event after the cutoff. Samples
whose event is already visible inside the window are excluded, since predicting it
is no longer early detection. A label $y_{i,q} = 0$ means only that the definition
is not met within the horizon, not that the token is safe.
Figure~\ref{fig:timeline} illustrates the setting.

\section{Methodology}
\label{sec:methodology}

\subsection{Data Collection}
\label{subsec:data}

\paragraph{Token selection.}
Token identifiers are taken from the 2024 portion of
SolRPDS~\citep{alhaidari2025}. We discard entries with an empty mint or pool
address, keep one entry per mint (27{,}752 mints), and draw a reproducible
random sample of 2{,}700 mints with a fixed seed. The SolRPDS mint field
occasionally holds the quote asset of a pool rather than the launched token; one
such asset in our sample, a stablecoin, has no liquidity history and never receives
a valid label. Most sampled tokens (2{,}541) were created in 2024; the final modeling
cohort spans September 2022 to October 2024.
From SolRPDS, we use only the mint addresses; labels, features, and transfers are
collected independently. The mint is the unit of analysis, because supply,
authorities, and balances are defined relative to it.

\paragraph{Data sources.}
Token-level Birdeye endpoints provide the creation time, one-minute liquidity
history and one-minute USD price candles over the first 24~hours for labeling,
and trades and liquidity snapshots within each window for the token-level features.
Transfers are recovered from Solana RPC (Section~\ref{subsec:graph}), which also
provides the creation time, as the block time of the mint's oldest transaction
signature. Intermediate label quantities are
stored only in the label table, and models read their inputs from an explicit list
of predictor columns.

\subsection{Operational Rug-Pull Labeling}
\label{subsec:labeling}

Following prior work that defines rug pulls by their observable market
outcome~\citep{mazorra2022donot,yaremus2025detecting,li2026catching}, we label
tokens with an explicit rule applied to Birdeye's one-minute liquidity history and
price candles within a nominal horizon $H = 24$~hours of creation, the period
in which most rug pulls occur~\citep{mazorra2022donot}. Because a liquidity drop alone can result from
benign events such as liquidity migration, and legitimate tokens can also crash in
price~\citep{mazorra2022donot,chen2026hypecollapse}, a token is a rug pull only if
three conditions hold jointly:
\begin{enumerate}
  \item \emph{Liquidity drop.} The drop time $\tau$ is the first snapshot at which
  liquidity falls more than $\theta_L = 80\%$ below the highest earlier level
  $L^{\mathrm{peak}}$; if no such snapshot exists, the token is negative.
  \item \emph{Price collapse.} The lowest price after $\tau$ is more than
  $\theta_P = 90\%$ below the highest price before $\tau$, using only
  candles entirely before or after $\tau$.
  \item \emph{Persistent depletion.} During the $\Delta = 30$~minutes after
  $\tau$, liquidity stays at or below $(1-\theta_L)\,L^{\mathrm{peak}}$ for at least
  $\rho_{\min} = 80\%$ of the time and is still below it at the end, which
  excludes liquidity that is removed and shortly re-added.
\end{enumerate}
$\theta_P$ follows the price-decline criterion of \citet{mazorra2022donot}, and
near-complete liquidity removal is a common rug-pull
indicator~\citep{cernera2023token,yaremus2025detecting}; $\Delta$ and $\rho_{\min}$
are operational choices. All thresholds were fixed before labeling. Varying
$\theta_P$ between 0.80 and 0.95 or $\rho_{\min}$ between 0.60 and
0.90 changes the number of positive labels by at most 3.6\%
(Appendix~\ref{app:labeling}, which also gives the formal definitions).

The label does not depend on the window; the window only determines whether a
sample is valid. If the drop is already visible inside the window ($\tau \le t_q$),
the sample is an \texttt{IN\_WINDOW\_EVENT} and excluded; a drop within
$\delta = 60$~s after the cutoff cannot be placed reliably at one-minute
resolution and is also excluded, and samples without sufficient liquidity or price history to determine
the label receive an \texttt{UNKNOWN} status (Appendix~\ref{app:labeling}). For every valid positive sample, all evidence of the rug pull therefore lies after
$t_q + \delta$. Because a negative label is confirmed only when the horizon ends,
every label becomes available at the end of the horizon, $t_0 + H$, which we use for
the embargo (Section~\ref{subsec:split}). A positive label is an operational
outcome, not proof of intent. Slower rug pulls and schemes without a liquidity
drain, such as freeze-authority abuse, are not captured; \citet{chen2026hypecollapse}
found nine such cases among 100 tokens not flagged within a comparable
24-hour window.

\subsection{Observation Windows}
\label{subsec:windows}

The three windows are the 25th, 50th, and 75th percentiles of the Solana rug-pull
lifespan distribution of \citet{chen2026hypecollapse}, measured on 76{,}469
rug-pull tokens from the first half of 2025: $w_{q_1} = 750$~s,
$w_{q_2} = 2{,}127$~s, and $w_{q_3} = 5{,}585$~s, i.e., about
12.5, 35.5, and 93~minutes. They were fixed before data collection,
since labels and features are defined relative to $t_q$, and we report all three.
On our data, the windows remain early decision points: of the 765 rug pulls
confirmed in $q_1$, only 15.3\% have their liquidity drop by the $q_2$ cutoff
and 38.0\% by the $q_3$ cutoff, so most occur after even the latest window.
Rug pulls that happen before the $q_1$ cutoff, or that lack sufficient data to be
labeled, cannot be counted, so these shares refer to confirmed rug pulls only.

As the window grows, more tokens leave the task as in-window events (128,
321, and 590), leaving 1{,}961, 1{,}798, and 1{,}573
\emph{samples}, i.e., valid token--window pairs with token-level features (one
token with a valid label has no stored feature row, so
Table~\ref{tab:label-status} lists 1{,}962, 1{,}799, and 1{,}574
valid labels). Because
the excluded tokens are mostly early rug pulls, the rug-pull rate falls from
39.0\% to 35.9\% and 30.1\%. The populations are nested, so a
difference between windows can reflect longer observation or the loss of the
fastest rug pulls; the panel scope (Section~\ref{subsec:protocol}) separates the two.

\subsection{Token-Level Features and Missing Data}
\label{subsec:features}

Each sample has five token-level features computed from Birdeye trades in
$[t_0, t_q)$ and cutoff liquidity snapshots (Table~\ref{tab:token-features}). Liquidity at the
cutoff is the latest snapshot at or before $t_q$, at most 60~s old; a snapshot stamped
exactly at $t_q$ is possible for at most 13, 18, and 14 samples.

\begin{table}[t]
  \caption{Missing token-level features by window (\%). Changes are percentage
  changes from the first to the second half-window; volumes and liquidity
  are measured in USD. Wallet counts include distinct addresses in buy
  and sell swaps.}
  \label{tab:token-features}
  \small
  \setlength{\tabcolsep}{3pt}
  \begin{tabular}{lrrr}
    \toprule
    Feature & $q_1$ & $q_2$ & $q_3$ \\
    \midrule
    Sell-volume change  & 66.8 & 51.2 & 45.3 \\
    Buy-volume change   & 65.4 & 50.8 & 45.3 \\
    Unique-wallet count & 5.0  & 6.3  & 15.8 \\
    Unique-wallet change & 65.4 & 50.8 & 45.2 \\
    Liquidity at cutoff & 57.9 & 41.1 & 34.5 \\
    \bottomrule
  \end{tabular}
\end{table}

\paragraph{Missing values.}
Missing features arise from undefined percentage changes when the first-half
baseline is zero, incomplete trade histories, or unavailable cutoff liquidity.
Missingness is associated with the outcome: complete cases have higher rug-pull
rates than the original cohort (Appendix~\ref{app:graph}). We impute missing
values with training medians and apply quantile transformations fitted on the
training partition. The main setting uses the original cohort without
missingness indicators. Ablations add binary missingness indicators and
restrict the cohort to samples with fewer than three missing features or
to complete cases.

\subsection{Transfer-Graph Construction}
\label{subsec:graph}

\paragraph{Transfer recovery and cleaning.}
For each mint, we retrieve the signatures of transactions that reference the mint
(\texttt{getSignatures\allowbreak{}For\allowbreak{}Address}), fetch the transactions, and parse the SPL Token
and Token-2022 transfer instructions, including inner instructions, recording
signature, block time, endpoints, and amount. Only transfers with a block time in
$[t_0, t_q)$ are kept, and records with missing fields, self-transfers, and
duplicates (same token, window, signature, source, destination, and amount) are
removed, which reduces 2{,}864{,}233 records to 2{,}862{,}579. Each
retained transfer is one directed edge with relative time
$\phi = (t - t_0)/w_q \in [0,1)$. Transfer amounts are not used, because raw token amounts are not comparable across
tokens, so the graph encodes topology and timing only.

\paragraph{Graph-bearing samples.}
A sample without retained transfers has no graph rather than an empty one, because
an absence of recovered transfers cannot be distinguished from incomplete recovery.
In the original cohort, 905, 950, and 884 samples
(46.4\%, 53.1\%, 56.4\%) are \emph{graph-bearing}, with rug-pull
rates of 43.6\%, 39.2\%, and 31.9\%. Among tokens valid in all
windows, every token with a graph in $q_1$ also has one later.

\paragraph{Features.}
Each endpoint has six features: its numbers of outgoing and incoming transfers and
of distinct receivers and senders ($\log(1+x)$), and the earliest and mean relative
time of its transfers. Each graph is summarized by nine features: the numbers of
transfers, endpoints $N$, distinct ordered pairs $M$, senders, and receivers, the
density $M/(N(N-1))$, the average degree $2M/N$, and the earliest and mean transfer
time (Appendix~\ref{app:graph}). Several are strongly dependent; for example,
endpoints and receivers have a Spearman correlation of 1.00 in every window.

\begin{figure}[t]
  \centering
  \includegraphics[width=\columnwidth]{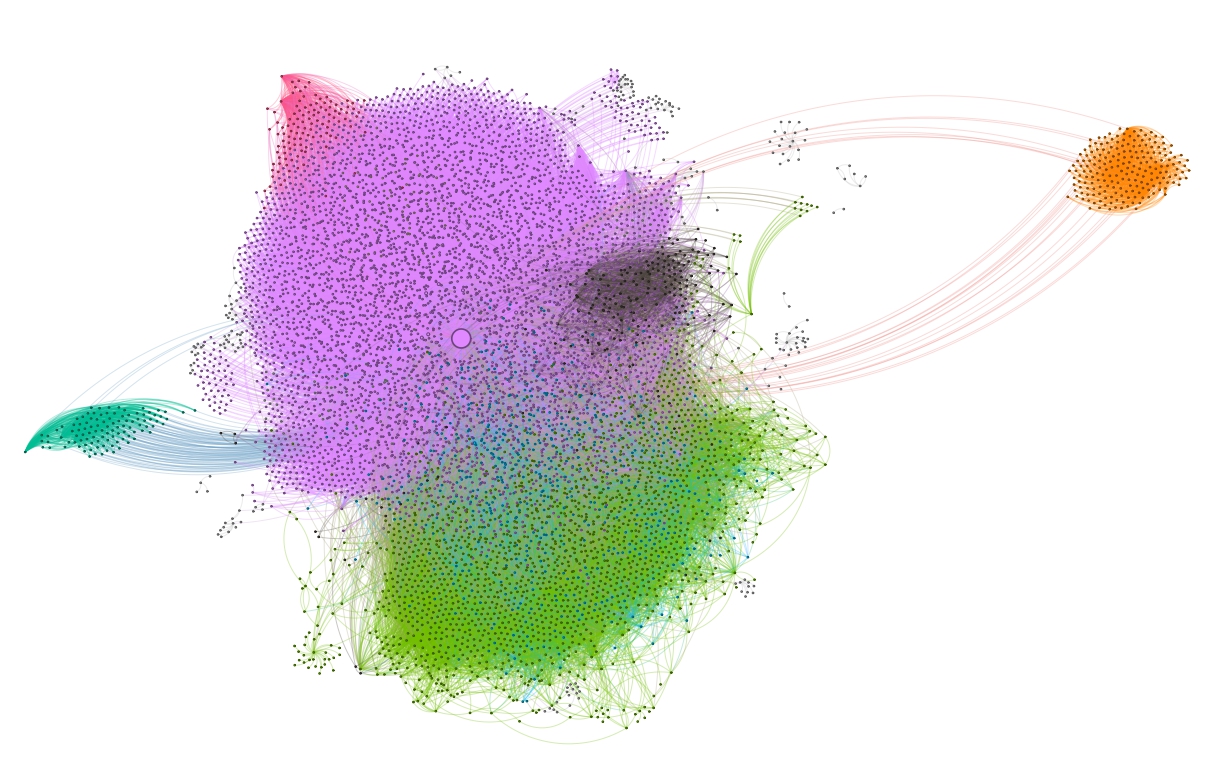}
  \caption{Recovered transfer network, visualized with Gephi.\protect\footnotemark{}
    Only the 8{,}026 transfer endpoints with at least five distinct neighbors
    are shown. Edges represent transfers, larger nodes indicate greater activity,
    and colors distinguish network communities rather than rug-pull labels.}
  \Description{A force-directed layout of a large transfer network with several
  dense colored communities of endpoints, a high-activity hub near the center, and
  a small separate community connected to the rest by a few long edges.}
  \label{fig:transfer-graph}
\end{figure}

Figure~\ref{fig:transfer-graph} shows the recovered transfers of all tokens as a
single network. Endpoints shared between tokens connect their graphs only in this
view; the models use a separate graph per token and window, and a single global
graph did not improve them (Section~\ref{subsec:ablations}).

\footnotetext{\url{https://gephi.org/}}

\subsection{Models}
\label{subsec:models}

\paragraph{Non-graph models.}
We evaluate a prior-probability classifier, logistic regression, a decision tree,
random forest, gradient boosting, XGBoost, a multilayer perceptron,
FT-Transformer~\citep{gorishniy2021revisiting}, and
AutoInt~\citep{song2019autoint}. Each is trained on four views: the five
token-level features on all samples (\emph{node}) or on graph-bearing samples
(\emph{node-matched}), the nine graph-summary features (\emph{graph}), and all
fourteen (\emph{node--graph}), the matched counterpart of the GNNs.

\paragraph{Graph neural networks.}
GCN~\citep{kipf2016semi}, GraphSAGE~\citep{hamilton2017inductive}, and
GAT~\citep{velivckovic2017graph} classify each graph from the six node features
with $L$ message-passing layers (ReLU, or ELU for GAT, and dropout). Messages follow
the transfer direction, so an endpoint aggregates from its senders; parallel edges are kept, and GCN and GAT also include each endpoint's own
representation when aggregating its neighbors. Mean, max, or sum pooling yields a graph
embedding $\mathbf{h}_G$, which is concatenated with the fourteen token-level and
graph-summary features $\mathbf{g}_{i,q}$:
$\hat{y}_{i,q} = \sigma(\operatorname{MLP}([\mathbf{h}_G \,\|\, \mathbf{g}_{i,q}]))$,
with a two-layer MLP. Both families thus receive the same fourteen features, and
the GNNs additionally learn from endpoint-level structure. Timing enters through
endpoint and graph summaries rather than explicit edge-time message passing.

\subsection{Training and Evaluation Protocol}
\label{subsec:protocol}

\paragraph{Chronological split and embargo.}
\label{subsec:split}
A deployed system must score newly created tokens using models trained on
historical data. We therefore evaluate on tokens created after those used
for training. A random split, including a stratified random split, would
mix earlier and later tokens across partitions, allowing models to learn
from tokens newer than those in the test set. Such temporal inconsistency
can inflate reported performance~\citep{pendlebury2019tesseract}.
A chronological split also exposes changes in the token population over
time (Section~\ref{subsec:splitgap}). We sort the 1{,}961 tokens
eligible at $q_1$ by creation time and assign the oldest 70\% to
training, the next 15\% to validation, and the newest 15\%
to testing, without using labels. Each token retains its partition
assignment across all windows, cohorts, and models. A token is embargoed, i.e.,
excluded, when its label becomes available ($t_0 + H$) only after the next
partition begins, so that no label is unknown at the time it would be used; this
removes 9, 8, and 7 tokens in $q_1$, $q_2$, and $q_3$.

\paragraph{Evaluation scopes.}
The \emph{full} scope uses all samples of a cohort in each window, the population
a deployed system faces at $t_q$; graph views and GNNs use its graph-bearing
samples, and models are always compared on identical test tokens. The \emph{panel}
scope keeps only tokens that are valid and graph-bearing in all windows, so that
differences between windows reflect observation time alone. It contains
704 tokens in the original cohort with rug-pull rate 34.8\%. The panel is selective: requiring validity in $q_3$
removes 386 $q_1$ tokens with a rug-pull rate of 74.9\%, so panel
results are a controlled comparison across windows, not an estimate for all tokens.

\paragraph{Training.}
The reference results use frozen ROC-AUC-selected runs; subsequent AP-selected
experiments are separate.
Hyperparameters are selected with a separate Optuna TPE search per model and
configuration~\citep{akiba2019optuna} (60 trials for classical models,
40 for neural and graph models), maximizing validation ROC-AUC. The GNN
search covers hidden size, depth (1--3), dropout, pooling, learning
rate, weight decay, batch size, attention heads, and positive-class weighting;
class weighting is also tuned for logistic regression, the decision tree, and the
random forest. Neural models stop after 25 epochs without improvement in
validation ROC-AUC (at most 200). Preprocessing is fitted on the training
partition, and each selected configuration is retrained with five seeds.

\paragraph{Metrics and statistical testing.}
The primary metric is Average Precision (AP); ROC-AUC is secondary. Our reported
comparisons use threshold-free metrics, since fixed thresholds transfer poorly under
the prevalence shift. Before evaluating on the test partition, we fixed one primary
analysis of earlier predictive performance: in the panel and original cohort without indicators,
each of GCN, GraphSAGE, and GAT at \emph{$q_1$} is compared with XGBoost on the
node--graph view at $q_1$, $q_2$, and $q_3$, i.e., with a non-graph model that has
observed the tokens for 12.5, 35.5, or 93~minutes. For each of the nine comparisons, we measure the difference in test AP
averaged over five seeds. We estimate uncertainty using 10{,}000
paired bootstrap samples: test tokens are resampled within each class,
and the same resampled tokens are used for both models. We apply a Bonferroni correction~\citep{bonferroni1936teoria} to obtain
confidence intervals with a joint confidence level of 95\%
across all nine comparisons.

All other comparisons are exploratory and use paired-bootstrap
95\% confidence intervals. For within-window comparisons, we also report
Holm-adjusted $p$-values~\citep{holm1979simple}. Model-variant comparisons are not
corrected for multiple testing; these variants
(Section~\ref{subsec:ablations}) were developed after the primary
analysis had been evaluated.

\section{Results}
\label{sec:results}

\subsection{Early Structural Differences}
\label{subsec:structure}

We first ask whether rug-pull and non-rug transfer graphs differ at all. We
compare the two classes in the training partition of the original cohort
(579, 632, and 599 graphs in $q_1$, $q_2$, $q_3$) with two-sided
Mann--Whitney $U$ tests~\citep{mann1947test} and Benjamini--Hochberg
correction~\citep{benjamini1995controlling} across all windows and
metrics, and report rank-biserial correlations $r$, with negative values indicating
higher values in non-rug graphs. This analysis did not guide the primary
feature or model choices.

Rug-pull and non-rug graphs already differ clearly after 12.5~minutes, in a
consistent pattern: non-rug activity concentrates around a few dominant receivers,
whereas rug-pull activity is spread over more endpoints with more one-directional
transfers. Rug-pull graphs involve more endpoints (median 145 vs.\
58.5 in $q_1$) but are sparser. The largest differences concern
receiver-side centralization: the maximum in-degree centrality and maximum PageRank
are clearly higher in non-rug graphs ($r = -0.42$ for both in $q_1$). Non-rug
graphs are also more reciprocal and have higher maximum betweenness. These
differences remain significant after correction in every window
(Appendix~\ref{app:structure}), whereas the sending side differs little: maximum
out-degree centrality differs only in $q_3$. This asymmetry between receiving and
sending endpoints motivates the direction-aware variants in
Section~\ref{subsec:ablations}.

\subsection{Earlier Detection Across Windows (RQ1, Primary Analysis)}
\label{subsec:primary}

RQ1 asks whether a graph model at the earliest cutoff can outperform a non-graph
model that has observed the same tokens for as long or longer.
Table~\ref{tab:primary} reports the pre-specified analysis on the panel
test set, where the prior-probability classifier reaches an AP of
0.279. None of the nine contrasts supports superiority: all nine Bonferroni-adjusted intervals contain zero and are wide enough to be
compatible with both AP losses and substantial gains. The larger differences at
later windows reflect only XGBoost's declining AP on the same tokens, since the
GNN predictions are fixed at $q_1$.

XGBoost was fixed as the reference before test evaluation and had the highest
validation ROC-AUC among non-graph models in the panel at $q_1$ and $q_2$. Three
exploratory checks assess whether the conclusion depends on this choice
(Appendix~\ref{app:rq1}). First, re-tuning XGBoost with a positive-class weight changes its AP by at most
0.028 and leaves all nine contrasts unsupported. Second, replacing XGBoost by
the non-graph model with the highest validation ROC-AUC in each window
(FT-Transformer at $q_3$) yields $\Delta$AP between -0.05 and +0.07,
with no supported contrast. Third, across all 72 combinations of graph model,
non-graph model, and window, only contrasts against the decision tree, the weakest
non-graph model, have intervals that exclude zero. We therefore find no
statistically supported AP advantage of a graph model at 12.5~minutes over a
competitive non-graph model observing the same tokens for 12.5, 35.5,
or 93~minutes.

\begin{table}[t]
    \caption{Primary analysis (panel, $n = 136$, 38 positives). Top: test
  AP (mean $\pm$ sd over five seeds). Bottom: $\Delta$AP of each GNN at $q_1$ minus
  XGBoost at each window, with Bonferroni-adjusted 95\% intervals.}
  \label{tab:primary}
  \small
  \setlength{\tabcolsep}{3pt}
  \begin{tabular}{lccc}
    \toprule
    & $q_1$ (12.5~min) & $q_2$ (35.5~min) & $q_3$ (93~min) \\
    \midrule
    \multicolumn{4}{l}{\emph{Test AP}} \\
    GCN       & .477\,$\pm$\,.036 & .473\,$\pm$\,.021 & .460\,$\pm$\,.019 \\
    GraphSAGE & .522\,$\pm$\,.011 & .528\,$\pm$\,.022 & .536\,$\pm$\,.009 \\
    GAT       & .457\,$\pm$\,.023 & .513\,$\pm$\,.014 & .481\,$\pm$\,.018 \\
    XGBoost   & .464\,$\pm$\,.015 & .452\,$\pm$\,.011 & .393\,$\pm$\,.006 \\
    \midrule
    \multicolumn{4}{l}{\emph{$\Delta$AP: GNN at $q_1$ minus XGBoost}} \\
    GCN       & +.013 {\scriptsize[-.122, .122]} & +.026 {\scriptsize[-.142, .173]} & +.084 {\scriptsize[-.089, .249]} \\
    GraphSAGE & +.058 {\scriptsize[-.119, .205]} & +.070 {\scriptsize[-.134, .251]} & +.129 {\scriptsize[-.062, .295]} \\
    GAT       & -.008 {\scriptsize[-.120, .090]} & +.005 {\scriptsize[-.150, .145]} & +.063 {\scriptsize[-.093, .212]} \\
    \bottomrule
  \end{tabular}
\end{table}

\subsection{Graph versus Non-Graph Models within Each Window (RQ2)}
\label{subsec:within}

RQ2 asks whether a graph model outperforms a non-graph model that has observed the
token for the same time. In each scope and window, we take the graph model (GCN,
GraphSAGE, or GAT) and the non-graph model (node--graph view) with the highest
validation ROC-AUC and compare them on the same test tokens (Table~\ref{tab:within}).
Unlike in the primary analysis, the compared models are chosen by validation
performance and may differ between cells, so this analysis is exploratory. Using seed-ensemble
predictions, the selected graph model achieves higher test AP in 5 of 6 cells; the
exception is panel $q_3$, where FT-Transformer performs better. ROC-AUC is similar
between the two families. Adding comparisons with XGBoost in the four cells where
it was not the selected non-graph model gives 10 comparisons. Three have
unadjusted intervals that exclude zero, but none remains significant after Holm
correction (smallest adjusted $p = 0.14$). The higher AP at similar ROC-AUC
is consistent with better ranking near the top of the prediction list, although the
evidence does not establish a reliable advantage. Brier scores improve over the
prior-probability classifier in some cells and worsen in others, providing no
consistent evidence of better probability estimates.

\begin{table}[t]
    \caption{Exploratory within-window comparison of the selected GNN and non-graph
  model: test AP / ROC-AUC of seed-ensemble predictions (scores averaged over five
  seeds, hence slightly different from the per-seed means in
  Tables~\ref{tab:primary} and~\ref{tab:split-gap}), and $\Delta$AP with unadjusted
  paired-bootstrap 95\% interval. Prev.: test prevalence.}
  \label{tab:within}
  \small
  \setlength{\tabcolsep}{2.5pt}
  \begin{tabular}{llllc}
    \toprule
    Cell & Prev. & GNN & Non-graph & $\Delta$AP [95\% CI] \\
    \midrule
    Panel $q_1$ & .279 & SAGE .521/.667 & XGB .464/.678 & +.058 {\scriptsize[-.075, .172]} \\
    Panel $q_2$ & .279 & GAT .520/.677 & XGB .460/.648 & +.060 {\scriptsize[-.041, .145]} \\
    Panel $q_3$ & .279 & GCN .489/.684 & FT-T .542/.687 & -.053 {\scriptsize[-.153, .054]} \\
    Full $q_1$  & .327 & GAT .635/.728 & LR .569/.728 & +.066 {\scriptsize[-.010, .125]} \\
    Full $q_2$  & .303 & SAGE .597/.726 & FT-T .489/.689 & +.108 {\scriptsize[.019, .184]} \\
    Full $q_3$  & .269 & GCN .533/.703 & MLP .491/.686 & +.042 {\scriptsize[-.020, .102]} \\
    \bottomrule
  \end{tabular}
\end{table}

\paragraph{Statistical power.}
Our test sets provide limited power to detect gains of the observed
size. Based on bootstrap standard errors and a normal approximation,
within-window GNN--XGBoost comparisons require AP differences of about
0.10--0.18 for 80\% power at $\alpha = 0.05$.
Multiple-testing correction raises these thresholds further.
Most observed gains fall below these thresholds, so the absence of
significant differences should not be read as evidence that the
two model families are equivalent.

\subsection{Training, Validation, and Test Performance}
\label{subsec:splitgap}

Validation scores are far higher than test scores for every model: AP falls from
0.82--0.91 on validation to 0.45--0.63 on test, and
ROC-AUC from 0.84--0.91 to 0.65--0.73
(Appendix~\ref{app:splitgap}). The evidence points mainly to a temporal shift
between the periods rather than to overfitting. The validation period has more rug
pulls, but re-weighting it to the test prevalence explains only
16--41\% of the AP gap and none of the ROC-AUC drop. The token
population also changes: pump.fun mints make up 12--14\% of training,
36--40\% of validation, and 63--67\% of test samples, and a
token's nearest training tokens predict its label well for validation tokens but
poorly for test tokens (transfer $k$NN AUC 0.77--0.79 vs.\
0.53--0.67). In contrast, selection effects are small: the best search
trial exceeds the median trial by at most 0.034 ROC-AUC. Validation scores
therefore overstate deployed performance. A random split would also overstate it: in a separate stratified split, test
ROC-AUC of XGBoost and GraphSAGE was 0.07--0.17 higher than under the
chronological split, although on a slightly different panel.

\subsection{Ablations and Model Variants}
\label{subsec:ablations}

\paragraph{Feature views.}
Adding the nine graph-summary features to the token-level features increases the
median test AP of the non-graph models by 0.029 in both evaluation scopes.
However, few configurations have unadjusted confidence intervals that exclude zero:
10 of 135 in the full scope, all favoring the added features, and
6 of 135 in the panel, of which one favors the token-level features
alone. On their own, graph-summary features perform about as well as token-level
features in the full scope (median $\Delta$AP $= +0.004$). In panel $q_1$, a token-only random forest even reaches the highest ROC-AUC of all
models (0.744), but a lower AP than the selected GNN.

GNNs using only the transfer graph achieve panel test AP of 0.507,
0.539, and 0.450 in $q_1$--$q_3$, compared with 0.522,
0.513, and 0.460 for the full models. The additional features therefore
provide no consistent improvement in these runs, although independent model
selection prevents attributing the differences solely to their removal.

\paragraph{Missingness.}
Adding missingness indicators changes the median test AP by less than 0.01
for both the non-graph models and the graph neural networks, with intervals that
exclude zero in at most 3 of 24 non-graph and 1 of 9 graph
configurations. Restricting training to the cohorts with fewer missing features does
not change the conclusions; most intervals include zero. 

\paragraph{Model variants.}
After the primary analysis, we evaluated several variants, each compared with
the selected graph neural network of the same cell on identical test tokens.
Motivated by the receiver--sender asymmetry of Section~\ref{subsec:structure},
direction-aware versions of GraphSAGE and GCN~\citep{rossi2024edge} aggregate
incoming and outgoing neighbors with separate weights. They were tuned on
validation AP rather than the validation ROC-AUC used for the reference models, so
their differences may partly reflect the selection criterion.
We detect no difference from the reference models ($\Delta$AP between -0.030
and +0.033, all intervals including zero). A single graph over all tokens, in which shared endpoints connect activity
before each target's cutoff (hub filtering uses corpus-wide memberships), was
evaluated only at $q_1$; it does not improve GraphSAGE or GCN in either scope
(ensemble $\Delta$AP between -0.060 and +0.022, all intervals including
zero). Averaging
GraphSAGE over five models trained on different folds of the training partition
improves ensemble AP in 4 of 6 cells, by up to 0.069, but no
interval excludes zero. Feeding the graph model's score into gradient boosting or XGBoost lowers AP
relative to the graph model in 11 of 12 comparisons. Finally, an adaptation of
HNN4RP~\citep{li2025hnn4rp} to Solana, without its smart-contract code embedding,
is worse than the reference model in every cell
(Appendix~\ref{app:variants}). None of these variants changes the conclusions of
Sections~\ref{subsec:primary} and~\ref{subsec:within}.

\section{Discussion and Limitations}
\label{sec:discussion}

\paragraph{Effect of longer observation.}
Within the panel, neither graph nor non-graph models improve
systematically when observation increases from 12.5 to
35.5 or 93~minutes, and XGBoost's AP decreases.
The panel includes only tokens that remain eligible for prediction
at $q_3$ and already have recovered transfers in $q_1$. This finding
therefore applies to tokens with early visible activity whose
rug-pull event, if any, occurs after the latest cutoff; it does not
extend to faster rug pulls excluded from the panel.
Limited power also prevents concluding that longer observation
adds no predictive information.

\paragraph{Possible reasons for modest graph gains.}
Several properties of the setting may limit the additional value of
a GNN. The recovered graphs are partial and mint-visible, and encode
topology and timing but no transfer amounts. Non-graph models already
receive structural information through graph-summary features, which
increase their median AP by 0.029. Graph availability also limits
coverage: only 46.4\% of samples at $q_1$ are graph-bearing.
With only 136--159 test tokens, the comparisons have limited
power to detect small AP differences (Section~\ref{subsec:within}).
The mostly positive seed-ensemble AP differences at similar ROC-AUC
are compatible with a benefit among the highest-ranked tokens, but
larger test sets are needed to establish whether this pattern is reliable.

\paragraph{Implications for Web3 platforms.}
For platforms that list new Solana tokens, our results suggest four
practical points. First, model scores computed from the first minutes
of activity can help prioritize tokens for review, but should not be
treated as reliable probabilities or definitive judgments.
Second, data availability is associated with the outcome: complete
cases and graph-bearing tokens have higher rug-pull rates
(Section~\ref{subsec:graph}, Appendix~\ref{app:graph}), and missingness
changes over time (Section~\ref{subsec:splitgap}). Platforms should
therefore monitor input coverage alongside model scores, even though
explicit missingness indicators provided little improvement in our
experiments. Third, the observed temporal shift (Section~\ref{subsec:splitgap}) motivates
ongoing performance monitoring and reassessment of training and calibration as the
token population changes. Fourth, claims of improved detection
from graph-based scoring should be supported by paired comparisons,
multiple seeds, appropriate multiple-testing correction, and test
sets large enough to detect practically meaningful gains.

\paragraph{Limitations.}
The label covers rug pulls that drain liquidity within 24~hours; slower rug
pulls and schemes without a liquidity drain are out of scope, and a positive label is an operational outcome rather than proof of
intent. The tokens come from one SolRPDS file, mostly from 2024, and the transfer
graphs are partial. The comparisons rely on a single chronological split with test
sets of 136--159 tokens. Class weighting was tuned for the graph models but not for all non-graph
models; re-tuning XGBoost with class weighting did not change the primary
conclusion. One candidate feature, liquidity at creation, was dropped after the split had been
fixed because it was missing for 94--97\% of samples. Finally, the model
variants of Section~\ref{subsec:ablations} were designed after the primary results
were known; their comparisons are exploratory and not corrected for multiple
testing.

\section{Conclusion}
\label{sec:conclusion}

We asked whether transfer graphs enable earlier rug-pull prediction for newly
launched Solana tokens than matched non-graph models. Rug-pull and non-rug graphs
already differ structurally within 12.5~minutes. However, our pre-specified
analysis finds no statistically supported AP advantage for GNNs at this cutoff over
XGBoost observing the same tokens at the same or later cutoffs (RQ1). Within each
window, graph models tend to achieve higher AP than non-graph models sharing the
token-level and graph-summary features, but no tested difference remains
significant after multiple-testing correction (RQ2). Longer observation does not
consistently improve performance on the fixed panel, although limited power
prevents concluding that later observations add no information. For Web3
platforms, these findings support using early model scores to prioritize review,
while leaving the additional benefit of graph-based scoring unresolved. 

\paragraph{Future work.}
First, larger evaluation sets are needed to establish whether a small graph
benefit exists. Our cohort is limited to 2{,}700 mints because each token
requires its full transaction history and minute-level market data; applying the
protocol to larger, more recent Solana
corpora~\citep{chen2026hypecollapse,hu2026memetrans} would increase statistical
power. Second, since a rug pull is executed at a specific moment, future work could
model the transaction stream continuously, using continuous-time temporal graph
models~\citep{wu2024tokenscout,wu2025rugscreener} and change-point methods, to
detect when an attacker begins to prepare it.

\section*{Data and Code Availability}
Code, token identifiers, labels, and token-level features are available at \url{https://github.com/Saba-Heydaridoost/rugpull_detector}, together with scripts for data collection, feature extraction, model training, and evaluation. Edge-level data, processed EDA artifacts, and the external SolRPDS dataset are not currently included in the repository. The remaining project artifacts and data needed for reproduction will be released in a future version of the repository, subject to applicable data-provider restrictions. SolRPDS can be obtained separately from its original source. These materials, together with the provided scripts, will enable reproduction of the reported results.

\section*{Ethics Statement}
We use only public blockchain records and market data, collected read-only; no
transactions were sent and no users were contacted. Addresses are used as
pseudonymous identifiers, and we make no attempt to link them to real-world
identities. Labels follow an operational definition: a positive label is not a legal
judgment or proof of fraud, and a negative label does not certify a token as safe.
Predictions contain errors and are intended as early-warning signals for review, not
as a basis for accusing projects, blacklisting addresses, or making investment
decisions.

\bibliographystyle{ACM-Reference-Format}
\bibliography{references}

@article{li2025hnn4rp,
    author  = {Li, Chenchen and Jin, Hai and Wu, Minrui and Xiao, Jiang},
    title   = {{HNN4RP}: Heterogeneous Graph Neural Network for Rug Pull Detection on Ethereum},
    journal = {Journal of Electronics \& Information Technology},
    volume  = {47},
    number  = {10},
    pages   = {3395--3409},
    year    = {2025},
    doi     = {10.11999/JEIT250160}
}

@inproceedings{alhaidari2025,
    author    = {Alhaidari, Abdulrahman and Kalal, Bhavani and Palanisamy, Balaji and Sural, Shamik},
    title     = {{SolRPDS}: A Dataset for Analyzing Rug Pulls in Solana Decentralized Finance},
    booktitle = {Proceedings of the Fifteenth ACM Conference on Data and Application Security and Privacy},
    pages     = {293--298},
    year      = {2025},
    publisher = {Association for Computing Machinery},
    doi       = {10.1145/3714393.3726487}
}

@article{wu2025rphunter,
    author    = {Wu, Hao and Wang, Haijun and Li, Shangwang and Wu, Yin and Fan, Ming and Jin, Wuxia and Liu, Ting},
    title     = {{RPHunter}: Unveiling Rug Pull Schemes in Crypto Token via Code-and-Transaction Fusion Analysis},
    journal   = {IEEE Transactions on Software Engineering},
    year      = {2026},
    publisher = {IEEE},
    doi       = {10.1109/TSE.2026.3716306}
}

@article{yaremus2025detecting,
    title={Detecting Rug Pulls in Decentralized Exchanges: Machine Learning Evidence from the TON Blockchain},
    author={Yaremus, Dmitry and Li, Jianghai and Kalacheva, Alisa and Vodolazov, Igor and Yanovich, Yury},
    journal={arXiv preprint arXiv:2509.01168},
    year={2025}
}

@inproceedings{li2025trust,
    title={Trust Dynamics and Bot-Driven Responses: An Approach to Rug Pulls in Solana Meme Coin Markets},
    author={Li, Yueyao and Yao, Nanjun and Huo, Yuhui and Cai, Wei},
    booktitle={Proceedings of the 17th ACM Web Science Conference 2025},
    pages={106--116},
    year={2025},
    doi = {10.1145/3717867.3717922}
}

@inproceedings {cernera2023token,
    author = {Federico Cernera and Massimo La Morgia and Alessandro Mei and Francesco Sassi},
    title = {Token Spammers, Rug Pulls, and Sniper Bots: An Analysis of the Ecosystem of Tokens in Ethereum and in the Binance Smart Chain ({{{{{BNB}}}}})},
    booktitle = {32nd USENIX Security Symposium (USENIX Security 23)},
    year = {2023},
    isbn = {978-1-939133-37-3},
    address = {Anaheim, CA},
    pages = {3349--3366},
    url = {https://www.usenix.org/conference/usenixsecurity23/presentation/cernera},
    publisher = {USENIX Association},
    month = aug
}

@inproceedings{wu2024tokenscout,
    author    = {Wu, Cong and Chen, Jing and Zhao, Ziming and He, Kun and Xu, Guowen and Wu, Yueming and Wang, Haijun and Li, Hongwei and Liu, Yang and Xiang, Yang},
    title     = {TokenScout: Early Detection of {Ethereum} Scam Tokens via Temporal Graph Learning},
    booktitle = {Proceedings of the 2024 on ACM SIGSAC Conference on Computer and Communications Security},
    pages     = {956--970},
    year      = {2024},
    publisher = {Association for Computing Machinery},
    doi       = {10.1145/3658644.3690234}
}

@article{kalacheva2025memecoins,
    author  = {Kalacheva, Alisa and Kuznetsov, Pavel and Vodolazov, Igor and Yanovich, Yury},
    title   = {Detecting Rug Pulls in Decentralized Exchanges: The Rise of Meme Coins},
    journal = {Blockchain: Research and Applications},
    volume  = {7},
    number  = {2},
    pages   = {100336},
    year    = {2026},
    publisher={Elsevier},
    doi     = {10.1016/j.bcra.2025.100336}
}

@misc{chen2026hypecollapse,
    title         = {From Hype to Collapse: Investigating Rug Pull Scams on {Solana}},
    author        = {Chen, Jiaxin and Li, Ziwei and Jiang, Zigui and He, Ruihong and Zhou, Yantong and Wu, Jiajing and Zheng, Zibin},
    journal       = {arXiv preprint arXiv:2603.24625},
    year          = {2026},
    eprint        = {2603.24625},
    archivePrefix = {arXiv},
    primaryClass  = {cs.CR}
}

@misc{hu2026memetrans,
    author        = {Hu, Sihao and Tekin, Selim Furkan and Xu, Yichang and Liu, Ling},
    title         = {{MemeTrans}: A Dataset for Detecting High-Risk Memecoin Launches on {Solana}},
    year          = {2026},
    eprint        = {2602.13480},
    archivePrefix = {arXiv},
    url           = {https://arxiv.org/abs/2602.13480}
}

@misc{shoaei2026tmrugpull,
    title         = {{TMRugPull}: A Temporally Sound Multimodal Dataset for Early RugPull Detection},
    author        = {Shoaei, Fatemeh and Pishdar, Mohammad and Bag-Mohammadi, Mozafar and Karami, Mojtaba and Lagaisse, Bert},
    year          = {2026},
    eprint        = {2602.21529},
    archivePrefix = {arXiv},
    primaryClass  = {cs.CR},
    url           = {https://arxiv.org/abs/2602.21529}
}

@Article{mazorra2022donot,
    AUTHOR = {Mazorra, Bruno and Adan, Victor and Daza, Vanesa},
    TITLE = {Do Not Rug on Me: Leveraging Machine Learning Techniques for Automated Scam Detection},
    JOURNAL = {Mathematics},
    VOLUME = {10},
    YEAR = {2022},
    NUMBER = {6},
    publisher={MDPI},
    ARTICLE-NUMBER = {949},
    URL = {https://www.mdpi.com/2227-7390/10/6/949},
    ISSN = {2227-7390},
    DOI = {10.3390/math10060949}
}

@article{lin2024crpwarner,
    title={Crpwarner: Warning the risk of contract-related rug pull in defi smart contracts},
    author={Lin, Zewei and Chen, Jiachi and Wu, Jiajing and Zhang, Weizhe and Wang, Yongjuan and Zheng, Zibin},
    journal={IEEE Transactions on Software Engineering},
    volume={50},
    number={6},
    pages={1534--1547},
    year={2024},
    publisher={IEEE},
    doi = {10.1109/TSE.2024.3392451}
}

@misc{solanaTokens2026,
    author       = {{Solana}},
    title        = {Assets on Solana},
    year         = {2026},
    howpublished = {\url{https://solana.com/docs/tokens}},
    note         = {Accessed October 26, 2026}
}

@misc{solanaCore2026,
    author       = {{Solana}},
    title        = {Programs on Solana},
    year         = {2026},
    howpublished = {\url{https://solana.com/docs/core/programs}},
    note         = {Accessed September 5, 2026}
}

@misc{ethereumAccounts2026,
    author       = {{Ethereum.org}},
    title        = {Ethereum Accounts},
    year         = {2026},
    howpublished = {\url{https://ethereum.org/developers/docs/accounts/}},
    note         = {Accessed September 5, 2026}
}

@misc{li2026catching,
    title         = {Catching the Rug: Early Prediction of Fraudulent Memecoins on Solana via Machine Learning},
    author        = {Li, Jianghai and Kuznetsov, Pavel and Yanovich, Yury and Nott-Whaley, Konstantin and Vodolazov, Igor},
    journal={arXiv preprint arXiv:2608.20271},
    year          = {2026},
    eprint        = {2608.20271},
    archivePrefix = {arXiv},
    primaryClass  = {cs.AI}
}

@article{wu2025rugscreener,
    author  = {Wu, Cong and Cao, Hangcheng and Chen, Jing and Yan, Xiyu and Xu, Guowen and Zhao, Ziming and Liu, Yang and Jiang, Hongbo},
    title   = {{RUGSCREENER}: Leveraging Temporal Graph Neural Network for Rugpull Detection in {DeFi}},
    journal = {IEEE Transactions on Information Forensics and Security},
    volume  = {20},
    pages   = {11120--11133},
    year    = {2025},
    publisher={IEEE},
    doi     = {10.1109/TIFS.2025.3620669}
}

@inproceedings{liu2021pcgnn,
    author    = {Liu, Yang and Ao, Xiang and Qin, Zidi and Chi, Jianfeng and Feng, Jinghua and Yang, Hao and He, Qing},
    title     = {Pick and Choose: A {GNN}-based Imbalanced Learning Approach for Fraud Detection},
    booktitle = {Proceedings of the Web Conference 2021},
    pages     = {3168--3177},
    year      = {2021},
    publisher = {Association for Computing Machinery},
    doi       = {10.1145/3442381.3449989}
}

@inproceedings{bian2020bigcn,
    author  = {Bian, Tian and Xiao, Xi and Xu, Tingyang and Zhao, Peilin and Huang, Wenbing and Rong, Yu and Huang, Junzhou},
    title   = {Rumor Detection on Social Media with Bi-Directional Graph Convolutional Networks},
    booktitle={Proceedings of the AAAI conference on artificial intelligence},
    volume  = {34},
    number  = {01},
    pages   = {549--556},
    year    = {2020},
    doi     = {10.1609/aaai.v34i01.5393}
}

@inproceedings{huynh2025serial,
    author    = {Huynh, Phuong Duy and Dau, Son Hoang and Huppert, Nicholas and Cervenjak, Joshua and Sun, Hoonie and Tran, Hong Yen and Li, Xiaodong and Viterbo, Emanuele},
    title     = {Serial Scammers and Attack of the Clones: How Scammers Coordinate Multiple Rug Pulls on Decentralized Exchanges},
    booktitle = {Proceedings of the ACM Web Conference 2025},
    pages     = {1016--1033},
    year      = {2025},
    publisher = {Association for Computing Machinery},
    doi       = {10.1145/3696410.3714919}
}

@inproceedings{gorishniy2021revisiting,
    author    = {Gorishniy, Yury and Rubachev, Ivan and Khrulkov, Valentin and Babenko, Artem},
    title     = {Revisiting Deep Learning Models for Tabular Data},
    booktitle = {Advances in Neural Information Processing Systems},
    volume    = {34},
    pages     = {18932--18943},
    year      = {2021},
    url       = {https://proceedings.neurips.cc/paper/2021/hash/9d86d83f925f2149e9edb0ac3b49229c-Abstract.html}
}

@inproceedings{song2019autoint,
  title={Autoint: Automatic feature interaction learning via self-attentive neural networks},
  author={Song, Weiping and Shi, Chence and Xiao, Zhiping and Duan, Zhijian and Xu, Yewen and Zhang, Ming and Tang, Jian},
  booktitle={Proceedings of the 28th ACM international conference on information and knowledge management},
  pages={1161--1170},
  year={2019}
}

@inproceedings{kipf2016semi,
    author    = {Kipf, Thomas N. and Welling, Max},
    title     = {Semi-Supervised Classification with Graph Convolutional Networks},
    booktitle = {International Conference on Learning Representations},
    year      = {2017},
    url       = {https://mlanthology.org/iclr/2017/kipf2017iclr-semi/}
}

@inproceedings{velivckovic2017graph,
    author    = {Veli{\v{c}}kovi{\'c}, Petar and Cucurull, Guillem and Casanova, Arantxa and Romero, Adriana and Li{\`o}, Pietro and Bengio, Yoshua},
    title     = {Graph Attention Networks},
    booktitle = {International Conference on Learning Representations},
    year      = {2018},
    url       = {https://mlanthology.org/iclr/2018/velickovic2018iclr-graph/}
}

@inproceedings{hamilton2017inductive,
    author    = {Hamilton, Will and Ying, Zhitao and Leskovec, Jure},
    title     = {Inductive Representation Learning on Large Graphs},
    booktitle = {Advances in Neural Information Processing Systems},
    volume    = {30},
    pages     = {1024--1034},
    year      = {2017},
    url       = {https://proceedings.neurips.cc/paper/2017/hash/5dd9db5e033da9c6fb5ba83c7a7ebea9-Abstract.html}
}

@inproceedings{akiba2019optuna,
  title={Optuna: A next-generation hyperparameter optimization framework},
  author={Akiba, Takuya and Sano, Shotaro and Yanase, Toshihiko and Ohta, Takeru and Koyama, Masanori},
  booktitle={Proceedings of the 25th ACM SIGKDD international conference on knowledge discovery \& data mining},
  pages={2623--2631},
  year={2019}
}

@inproceedings{rossi2024edge,
    author    = {Rossi, Emanuele and Charpentier, Bertrand and Di Giovanni, Francesco and Frasca, Fabrizio and G{\"u}nnemann, Stephan and Bronstein, Michael M.},
    title     = {Edge Directionality Improves Learning on Heterophilic Graphs},
    booktitle = {Proceedings of the Second Learning on Graphs Conference},
    series    = {Proceedings of Machine Learning Research},
    volume    = {231},
    pages     = {25:1--25:27},
    year      = {2024},
    publisher = {PMLR},
    url       = {https://proceedings.mlr.press/v231/rossi24a.html}
}

@inproceedings{pendlebury2019tesseract,
  title={$\{$TESSERACT$\}$: Eliminating experimental bias in malware classification across space and time},
  author={Pendlebury, Feargus and Pierazzi, Fabio and Jordaney, Roberto and Kinder, Johannes and Cavallaro, Lorenzo},
  booktitle={28th USENIX security symposium (USENIX Security 19)},
  pages={729--746},
  year={2019}
}

@article{bonferroni1936teoria,
    author  = {Bonferroni, Carlo Emilio},
    title   = {Teoria Statistica delle Classi e Calcolo delle Probabilita},
    journal = {Pubblicazioni del R. Istituto Superiore di Scienze Economiche e Commerciali di Firenze},
    volume  = {8},
    pages   = {3--62},
    year    = {1936}
}

@article{holm1979simple,
  title={A simple sequentially rejective multiple test procedure},
  author={Holm, Sture},
  journal={Scandinavian journal of statistics},
  pages={65--70},
  year={1979},
  publisher={JSTOR}
}

@article{mann1947test,
    author  = {Mann, Henry B. and Whitney, Donald R.},
    title   = {On a Test of Whether One of Two Random Variables Is Stochastically Larger than the Other},
    journal = {The Annals of Mathematical Statistics},
    volume  = {18},
    number  = {1},
    pages   = {50--60},
    year    = {1947},
    doi     = {10.1214/aoms/1177730491}
}

@article{benjamini1995controlling,
  title={Controlling the false discovery rate: a practical and powerful approach to multiple testing},
  author={Benjamini, Yoav and Hochberg, Yosef},
  journal={Journal of the Royal statistical society: series B (Methodological)},
  volume={57},
  number={1},
  pages={289--300},
  year={1995},
  publisher={Wiley Online Library}
}

\appendix
\renewcommand{\dbltopfraction}{0.95}
\renewcommand{\dblfloatpagefraction}{0.5}
\setcounter{dbltopnumber}{4}
\setcounter{totalnumber}{6}

\section{Formal Definition of the Rug-Pull Label}
\label{app:labeling}

\begin{table*}[!t]
  \caption{Label statuses and their counts per observation window. Only
  \texttt{VALID} samples are used for supervised modeling.}
  \label{tab:label-status}
  \begin{tabular}{p{0.36\textwidth}p{0.42\textwidth}rrr}
    \toprule
    Status & Condition & $q_1$ & $q_2$ & $q_3$ \\
    \midrule
    \texttt{VALID} & Binary label determined from sufficient post-cutoff evidence & 1{,}962 & 1{,}799 & 1{,}574 \\
    \texttt{IN\_WINDOW\_EVENT} & First qualifying liquidity drop at or before the cutoff ($\tau \le t_q$) & 128 & 321 & 590 \\
    \texttt{UNKNOWN\_OBSERVATION\_BOUNDARY\_AMBIGUOUS} & Liquidity drop within $\delta$ after the cutoff & 4 & 6 & 1 \\
    \texttt{UNKNOWN\_MISSING\_LIQUIDITY\_HISTORY} & No liquidity snapshot within the horizon & 429 & 429 & 429 \\
    \texttt{UNKNOWN\_INSUFFICIENT\_LIQUIDITY\_HISTORY} & Fewer than two liquidity snapshots & 14 & 14 & 14 \\
    \texttt{UNKNOWN\_INVALID\_LIQUIDITY\_BASELINE} & No positive liquidity snapshot to serve as baseline & 14 & 14 & 14 \\
    \texttt{UNKNOWN\_INSUFFICIENT\_PERSISTENCE\_HORIZON} & Confirmation interval extends beyond $t_0 + H$ & 13 & 13 & 13 \\
    \texttt{UNKNOWN\_INSUFFICIENT\_PERSISTENCE\_COVERAGE} & Liquidity history ends before $\tau + \Delta$ & 41 & 36 & 30 \\
    \texttt{UNKNOWN\_MISSING\_PRICE\_HISTORY} & No price candle within the horizon & 4 & 3 & 3 \\
    \texttt{UNKNOWN\_MISSING\_PRE\_EVENT\_PRICE} & No complete price candle before $\tau$ & 52 & 29 & 11 \\
    \texttt{UNKNOWN\_MISSING\_POST\_EVENT\_PRICE} & No price candle after $\tau$ within the horizon & 39 & 36 & 21 \\
    \bottomrule
  \end{tabular}
\end{table*}

This appendix formalizes the three conditions of Section~\ref{subsec:labeling} and
reports the threshold sensitivity and label statuses.

\paragraph{Threshold sensitivity.}
Table~\ref{tab:label-sensitivity} varies $\theta_P$ or $\rho_{\min}$ with all other
parameters fixed, using the recorded ratios without recomputing event times.

\begin{table}[!htb]
  \caption{Number of positive labels under alternative price and persistence
  thresholds. The chosen values are $\theta_P = 0.90$ and
  $\rho_{\min} = 0.80$.}
  \label{tab:label-sensitivity}
  \small
  \begin{tabular}{lrrr}
    \toprule
    Setting & $q_1$ & $q_2$ & $q_3$ \\
    \midrule
    Chosen thresholds            & 765 & 645 & 474 \\
    $\theta_P = 0.80$      & 786 & 657 & 484 \\
    $\theta_P = 0.95$      & 753 & 637 & 470 \\
    $\rho_{\min} = 0.60$   & 784 & 662 & 491 \\
    $\rho_{\min} = 0.90$   & 746 & 628 & 461 \\
    \bottomrule
  \end{tabular}
\end{table}

\paragraph{Notation.}
The liquidity history returned by Birdeye is a chronologically ordered list of
snapshots. We index them by $k = 1, \ldots, n$ and write $s_k$ for the timestamp of
the $k$-th snapshot and $L_k$ for the USD liquidity it reports; only snapshots with
$t_0 \le s_k \le t_0 + H$ are used. We write $s_k$ rather than $t_k$ to distinguish
snapshot times from the reference times $t_0$ and $t_q$. Liquidity at an arbitrary
time $t$ is obtained by carrying the most recent snapshot forward:
$\tilde{L}(t) = L_k$ for $s_k \le t < s_{k+1}$. No synthetic observations are
inserted where the API returns none.

\paragraph{Condition 1: Liquidity drop.}
For every snapshot $k$, the liquidity drawdown relative to the running peak of all
earlier snapshots is
\[
D^{L}_k = 1 - \frac{L_k}{\max_{j<k} L_j} .
\]
The peak is taken only over earlier snapshots, so the drawdown never depends on
future observations; the first snapshot with positive liquidity serves only as the
initial baseline. The liquidity-drop time is the earliest crossing of the
threshold,
\[
\tau = \min \{\, s_k : D^{L}_k > \theta_L \,\} = s_{k^\ast},
\qquad
L^{\mathrm{peak}} = \max_{j<k^\ast} L_j .
\]
The scan stops at this first crossing. If no snapshot crosses the threshold within
the horizon, the token is negative and the remaining conditions are not evaluated.

\paragraph{Condition 2: Price collapse.}
Price is given as one-minute OHLCV candles indexed by $c$; candle $c$ starts at
$s_c$, covers $[s_c, s_c + 60)$ seconds, and has high $h_c$ and low $l_c$. The
pre-drop peak and post-drop minimum are
\[
P^{\mathrm{peak}} = \max_{\substack{s_c \ge t_0 \\ s_c + 60 \le \tau}} h_c ,
\qquad
P^{\min} = \min_{\lceil \tau \rceil_{60} \le s_c < t_0 + H} l_c ,
\]
where $\lceil \tau \rceil_{60}$ is $\tau$ rounded up to the next full minute.
Candles are included by their start time, so the final candle may extend
less than one minute beyond the nominal horizon.
A candle that straddles $\tau$ is used on neither side, so no post-drop
price enters the peak. The relative price decline is
$D^{P} = 1 - P^{\min} / P^{\mathrm{peak}}$, and the condition holds if
$D^{P} > \theta_P$. Outlier candles are retained, because the rule is defined on
the raw extrema.

\paragraph{Condition 3: Persistent depletion.}
With $\lambda = (1 - \theta_L)\, L^{\mathrm{peak}}$, the persistence ratio is the
fraction of the confirmation interval during which liquidity stays at or below
$\lambda$:
\[
\rho = \frac{1}{\Delta} \int_{\tau}^{\tau + \Delta}
\mathds{1}\!\left[ \tilde{L}(t) \le \lambda \right] dt ,
\]
where $\mathds{1}[\cdot]$ equals 1 when its argument is true and 0 otherwise.
Because $\tilde{L}(t)$ is piecewise constant, the integral is computed exactly as a
sum of segment lengths. The condition holds if $\rho \ge \rho_{\min}$ and
$\tilde{L}(\tau + \Delta) \le \lambda$. The confirmation interval must lie within
the outcome horizon and be covered by the liquidity history; otherwise persistence
cannot be evaluated.

\paragraph{Label.}
Combining the three conditions,
\[
y_{i,q} =
\mathds{1}\!\left[
\tau \text{ exists}
\;\wedge\; D^{P} > \theta_P
\;\wedge\; \rho \ge \rho_{\min}
\;\wedge\; \tilde{L}(\tau + \Delta) \le \lambda
\right],
\]
where $\wedge$ denotes logical conjunction.

\paragraph{Temporal separation of label evidence and predictors.}
The labeling histories span $[t_0, t_0 + H]$, so the reference levels
$L^{\mathrm{peak}}$ and $P^{\mathrm{peak}}$ may use observations inside the window,
but only as baselines for a later collapse. For every valid positive sample, all
evidence of the rug pull lies after the cutoff: $\tau > t_q + \delta$, $P^{\min}$
comes from candles starting at or after $\tau$, and the persistence interval starts
at $\tau$. For a valid negative sample, either no qualifying drop occurs within the
horizon, or the first drop occurs after $t_q + \delta$ but fails the price or
persistence condition. Predictors are computed from $[t_0, t_q)$, except for cutoff
liquidity, which may be observed at $t_q$, and none of the quantities above is used
as a feature or graph edge, following the leakage-aware design
of~\citet{shoaei2026tmrugpull}.

\paragraph{Label statuses.}
Table~\ref{tab:label-status} lists all statuses and their counts. Transient API or
RPC failures are retried rather than recorded (Section~\ref{subsec:data}), so the
\texttt{UNKNOWN} statuses reflect genuine data unavailability. The labeling
procedure defines further safeguard statuses, for example for events coinciding
with the end of the horizon; none of them occurred, and the statuses in
Table~\ref{tab:label-status} account for all 2{,}700 samples in every window.

\section{Cohorts and Graph Features}
\label{app:graph}

Table~\ref{tab:cohorts} gives the sizes and rug-pull rates of the three missingness
cohorts (Section~\ref{subsec:features}), and Table~\ref{tab:graph-features} defines the
endpoint-level and graph-level features (Section~\ref{subsec:graph}).

\begin{table}[!ht]
  \caption{Sizes and rug-pull rates of the missingness cohorts. The original cohort
  excludes the 9, 8, and 7 embargoed tokens.}
  \label{tab:cohorts}
  \small
  \setlength{\tabcolsep}{3.5pt}
  \begin{tabular}{lrrrrrr}
    \toprule
    & \multicolumn{2}{c}{$q_1$} & \multicolumn{2}{c}{$q_2$} & \multicolumn{2}{c}{$q_3$} \\
    \cmidrule(lr){2-3} \cmidrule(lr){4-5} \cmidrule(lr){6-7}
    Cohort & $n$ & Rug\,\% & $n$ & Rug\,\% & $n$ & Rug\,\% \\
    \midrule
    Original       & 1{,}952 & 39.0 & 1{,}790 & 35.9 & 1{,}566 & 30.2 \\
    $<3$ missing   & 676     & 58.1 & 883     & 50.1 & 860     & 33.1 \\
    Complete case  & 527     & 65.7 & 765     & 54.9 & 692     & 36.8 \\
    \bottomrule
  \end{tabular}
\end{table}

\begin{table}[!ht]
  \caption{Features derived from the transfer graph of a sample. Times are relative
  times $\phi$ within the window.}
  \label{tab:graph-features}
  \small
  \begin{tabular}{p{0.30\columnwidth}p{0.60\columnwidth}}
    \toprule
    Feature & Description \\
    \midrule
    \multicolumn{2}{l}{\emph{Node level (input of graph neural networks)}} \\
    Out-transfers   & $\log(1+x)$ of the number of outgoing transfers \\
    In-transfers    & $\log(1+x)$ of the number of incoming transfers \\
    Out-neighbors   & $\log(1+x)$ of the number of distinct receivers \\
    In-neighbors    & $\log(1+x)$ of the number of distinct senders \\
    First time      & Earliest time of an incident transfer \\
    Mean time       & Mean time of incident transfers \\
    \midrule
    \multicolumn{2}{l}{\emph{Graph level (summary features)}} \\
    Transfers       & Number of transfers (edges of the multigraph) \\
    Endpoints       & Number of endpoints $N$ \\
    Edges           & Number of distinct ordered endpoint pairs $M$ \\
    Senders         & Number of distinct sending endpoints \\
    Receivers       & Number of distinct receiving endpoints \\
    Density         & $M / \bigl(N(N-1)\bigr)$ \\
    Average degree  & $2M / N$ \\
    First time      & Time of the earliest transfer \\
    Mean time       & Mean time of all transfers \\
    \bottomrule
  \end{tabular}
\end{table}

\section{Structural Differences}
\label{app:structure}
Table~\ref{tab:structure} reports the rank-biserial correlations discussed in
Section~\ref{subsec:structure}.
\begin{table}[t]
  \caption{Structural differences between rug-pull and non-rug transfer graphs in
  the training partition: rank-biserial correlation $r$ (positive: larger for
  rug-pull graphs). All values are significant after Benjamini--Hochberg correction
  unless marked n.s.}
  \label{tab:structure}
  \small
  \begin{tabular}{lrrr}
    \toprule
    Metric & $q_1$ & $q_2$ & $q_3$ \\
    \midrule
    Endpoints                   & 0.24  & 0.17  & 0.19  \\
    Density                     & -0.27 & -0.21 & -0.22 \\
    Max.\ in-degree centrality  & -0.42 & -0.41 & -0.36 \\
    Max.\ PageRank              & -0.42 & -0.41 & -0.37 \\
    Reciprocity                 & -0.26 & -0.25 & -0.26 \\
    Max.\ betweenness           & -0.26 & -0.30 & -0.34 \\
    Max.\ out-degree centrality & n.s.        & n.s.        & 0.12  \\
    \bottomrule
  \end{tabular}
\end{table}

\section{Model Variants}
\label{app:variants}

\begin{table*}[!tp]
  \caption{Test AP of the model variants (mean over five seeds) and $\Delta$AP
  relative to the selected graph neural network of the same cell, with paired
  bootstrap 95\% intervals of seed-ensemble scores (unadjusted). Because AP is a
  per-seed mean and $\Delta$AP uses seed-ensemble scores, the two can differ in
  sign.}
  \label{tab:variants}
  \footnotesize
  \setlength{\tabcolsep}{3pt}
  \begin{tabular}{lllll}
    \toprule
    Cell (reference AP) & DirSAGE & DirGCN & Bagged GraphSAGE & HNN4RP \\
    \midrule
    Panel $q_1$ (GraphSAGE .522) & .521; +.006 [-.050, .064] & .509; +.002 [-.059, .061] & .527; +.004 [-.043, .043] & .492; -.020 [-.109, .077] \\
    Panel $q_2$ (GAT .513)       & .535; +.033 [-.017, .082] & .537; +.003 [-.059, .073] & .526; -.002 [-.064, .056] & .461; -.056 [-.130, .024] \\
    Panel $q_3$ (GCN .460)       & .492; +.030 [-.028, .089] & .471; -.026 [-.079, .050] & .558; +.069 [-.028, .157] & .454; -.055 [-.124, .042] \\
    Full $q_1$ (GAT .626)        & .620; +.020 [-.032, .073] & .636; +.021 [-.027, .073] & .641; +.006 [-.051, .068] & .540; -.087 [-.158, -.0004] \\
    Full $q_2$ (GraphSAGE .596)  & .580; -.002 [-.071, .067] & .600; +.016 [-.047, .078] & .592; -.007 [-.046, .024] & .510; -.089 [-.161, -.009] \\
    Full $q_3$ (GCN .493)        & .484; -.030 [-.089, .028] & .489; -.011 [-.066, .041] & .555; +.022 [-.040, .089] & .465; -.013 [-.084, .060] \\
    \bottomrule
  \end{tabular}
\end{table*}

All variants were developed after the primary analysis and use the original cohort
without missingness indicators. \emph{DirSAGE} and \emph{DirGCN} aggregate incoming
and outgoing neighbors separately (DirGCN mixes them with a tuned weight
$\alpha \in \{0.25, 0.5, 0.75\}$), use mean--max pooling and the
same late fusion, and are tuned on validation AP (40 trials). \emph{Bagged
GraphSAGE} averages five GraphSAGE models trained on four of five stratified folds
of the training partition. \emph{Stacking} adds the out-of-fold GraphSAGE score as a
feature to gradient boosting or XGBoost. The \emph{global graph} links per-token
graphs through shared endpoints, using only transfers before each token's cutoff;
its hub filter uses corpus-wide token memberships. Our \emph{HNN4RP} adaptation
samples time-respecting neighbors in a token--wallet graph and encodes them with an
attention-based BiLSTM, without the original's contract-code embedding or a
hyperparameter search.

\section{Additional Results}
\label{app:rq1}

\paragraph{Sensitivity of the primary analysis.}
Table~\ref{tab:rq1-grid} repeats the comparison of Section~\ref{subsec:primary}
against every non-graph model (node--graph view) of the panel, using the same
estimator: the difference in mean per-seed test AP of a graph model at $q_1$ and a
non-graph model at $q_1$, $q_2$, or $q_3$, with 10{,}000 paired bootstrap
samples of the 136 test tokens. Treated as a separate exploratory family, the
nine contrasts against the non-graph model with the highest validation ROC-AUC in
each window (underlined) have Bonferroni-adjusted intervals that all contain zero.
Across the 72 contrasts, unadjusted 95\% intervals exclude zero only
against the decision tree (bold); two of these remain significant after Holm
correction over the 72 contrasts.

\begin{table*}[!tp]
  \caption{Graph model at $q_1$ minus each non-graph model at $q_1$, $q_2$, or $q_3$:
  $\Delta$AP in mean per-seed test AP (panel, original cohort, no missingness
  indicators, $n = 136$). AP: mean per-seed test AP of the non-graph model.
  Underlined: non-graph model with the highest validation ROC-AUC in that window.
  Bold: unadjusted 95\% paired-bootstrap interval excludes zero.}
  \label{tab:rq1-grid}
  \footnotesize
  \setlength{\tabcolsep}{3pt}
  \begin{tabular}{lcccccccccccc}
    \toprule
    & \multicolumn{3}{c}{Non-graph AP} & \multicolumn{3}{c}{GCN at $q_1$} &
      \multicolumn{3}{c}{GraphSAGE at $q_1$} & \multicolumn{3}{c}{GAT at $q_1$} \\
    \cmidrule(lr){2-4} \cmidrule(lr){5-7} \cmidrule(lr){8-10} \cmidrule(lr){11-13}
    Non-graph model & $q_1$ & $q_2$ & $q_3$ & $q_1$ & $q_2$ & $q_3$ & $q_1$ & $q_2$ & $q_3$ & $q_1$ & $q_2$ & $q_3$ \\
    \midrule
    Logistic regression & .448 & .424 & .435 & +.030 & +.053 & +.043 & +.075 & +.098 & +.088 & +.009 & +.032 & +.022 \\
    Decision tree & .352 & .388 & .414 & \textbf{+.125} & \textbf{+.090} & +.064 & \textbf{+.170} & \textbf{+.134} & +.108 & \textbf{+.104} & +.069 & +.043 \\
    Random forest & .477 & .429 & .484 & +.000 & +.049 & -.006 & +.045 & +.093 & +.038 & -.021 & +.028 & -.027 \\
    Gradient boosting & .493 & .420 & .403 & -.016 & +.058 & +.074 & +.029 & +.102 & +.119 & -.037 & +.037 & +.054 \\
    XGBoost & .464 & .452 & .393 & \underline{+.013} & \underline{+.026} & +.084 & \underline{+.058} & \underline{+.070} & +.129 & \underline{-.008} & \underline{+.005} & +.063 \\
    MLP & .474 & .417 & .496 & +.003 & +.061 & -.019 & +.048 & +.106 & +.026 & -.017 & +.040 & -.040 \\
    FT-Transformer & .480 & .404 & .508 & -.002 & +.073 & \underline{-.030} & +.042 & +.118 & \underline{+.015} & -.023 & +.052 & \underline{-.051} \\
    AutoInt & .495 & .433 & .508 & -.018 & +.045 & -.030 & +.027 & +.090 & +.015 & -.038 & +.024 & -.051 \\
    \bottomrule
  \end{tabular}
\end{table*}

\subsection{Training, Validation, and Test Performance}
\label{app:splitgap}

\begin{figure*}[!tp]
  \centering
  \includegraphics[width=\textwidth]{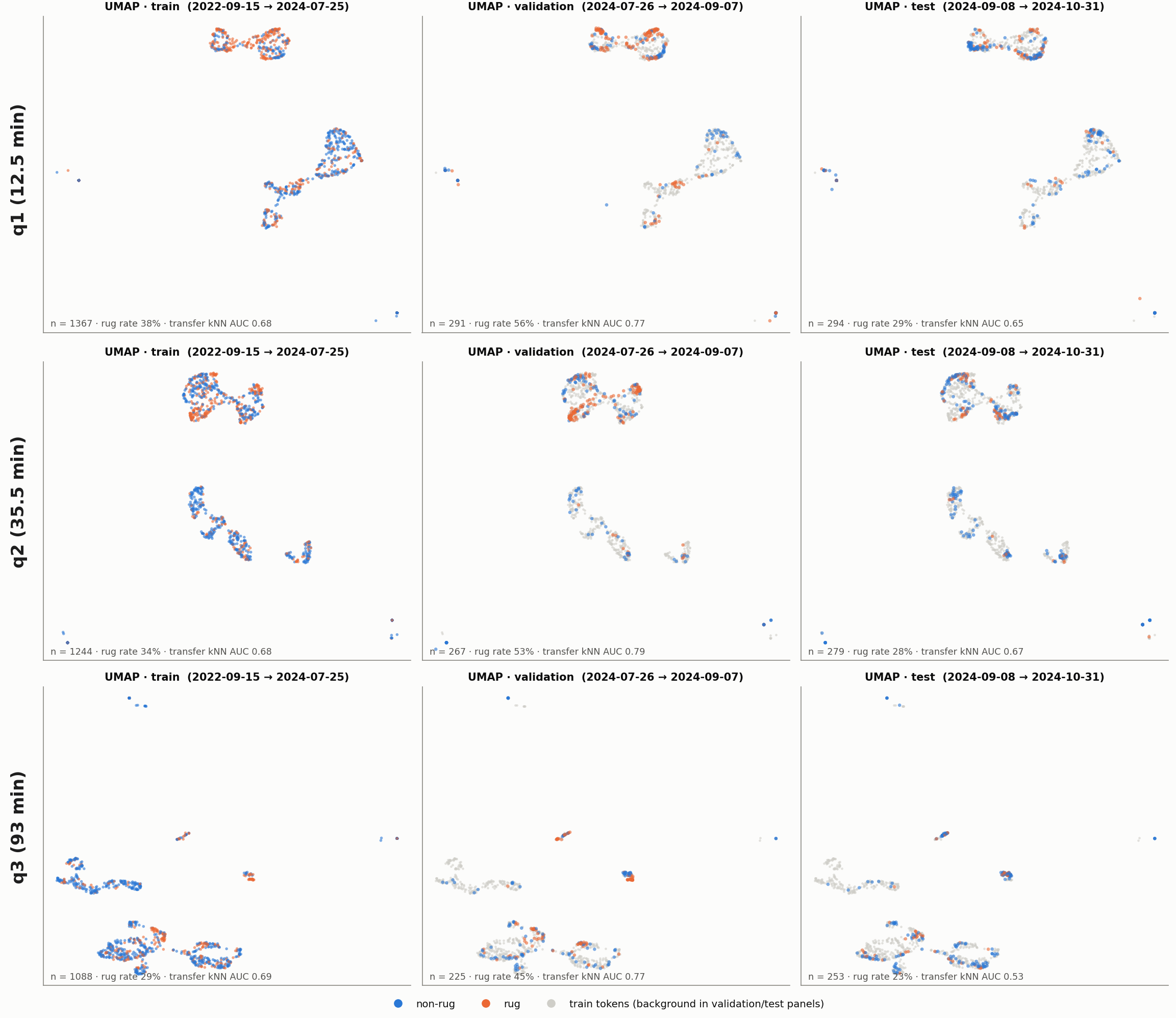}
  \caption{UMAP embedding of the samples of each window, split chronologically into
  training, validation, and test tokens (left to right; rows $q_1$--$q_3$). The
  embedding uses the token-level and graph-summary features with missingness
  indicators, with imputation, scaling, and UMAP fitted on training tokens only;
  validation and test tokens are projected through the fitted transforms and shown
  over the training tokens (grey). Transfer $k$NN AUC: AUC of the rug-pull share
  among each token's 15 nearest training tokens in the embedding (for
  training tokens, excluding the token itself).}
  \Description{Nine scatter plots in three rows (12.5, 35.5, and 93 minutes) and three
  columns (training, validation, test). Points are colored as rug pull or non-rug. In
  each row, validation tokens fall mostly into the rug-rich clusters of the training
  embedding, whereas test tokens fall into clusters that are mostly non-rug or mixed.}
  \label{fig:umap-split}
\end{figure*}

Table~\ref{tab:split-gap} reports the models of Table~\ref{tab:within} on all
three partitions, and Figure~\ref{fig:umap-split} shows the token population of
each partition (Section~\ref{subsec:splitgap}).

\paragraph{Prevalence.}
AP rises with the share of positives, and the validation period contains far more
rug pulls than the test period. Re-weighting the validation positives to the test
prevalence lowers validation AP by 0.07--0.12, which explains only
16--41\% of the gap. The remaining 0.17--0.37~AP, and the
drop in ROC-AUC, which does not depend on prevalence, must have other causes.

\paragraph{Population shift.}
The share of pump.fun mints rises from 12--14\% of training samples to
36--40\% of validation and 63--67\% of test samples, and data
availability changes with it: in $q_1$, 58\%, 33\%, and 43\% of
token-level predictor values are missing, and 42\%, 57\%, and
54\% of samples are graph-bearing in the three partitions. In a UMAP
embedding fitted on training tokens only (Figure~\ref{fig:umap-split}), validation
tokens concentrate in regions that were already rug-rich in training (transfer
$k$NN AUC 0.77, 0.79, and 0.77 in $q_1$--$q_3$), whereas test
tokens fall into regions that are mostly non-rug or mixed (0.65, 0.67,
and 0.53).

\paragraph{Validation is easier than training.}
For the graph models and logistic regression, validation ROC-AUC is at least as
high as training ROC-AUC (for example, 0.853 vs.\ 0.746 for GraphSAGE in
panel $q_1$), and the training neighbors of a training token are less predictive
(transfer $k$NN AUC 0.68--0.69) than those of a validation token. The
flexible tabular models fit the training data more closely (training ROC-AUC up to
0.92), but their validation scores are similar.

\paragraph{Selection on validation.}
The best of the 40--60 search trials exceeds the median trial by at
most 0.034 ROC-AUC, and the five retrained seeds stay within 0.02 of the
best trial, far less than the validation--test gap of 0.16--0.23.

\begin{table*}[!tp]
  \caption{Training, validation, and test performance of the models of
  Table~\ref{tab:within} (mean over five seeds). Prev.: rug-pull rate in the
  training/validation/test partitions (\%). Val@test: validation AP with positives
  re-weighted to the test prevalence. Training scores are those of the fitted
  models on their own training partition.}
  \label{tab:split-gap}
  \footnotesize
  \setlength{\tabcolsep}{4pt}
  \begin{tabular}{llcccccccc}
    \toprule
    & & & \multicolumn{4}{c}{AP} & \multicolumn{3}{c}{ROC-AUC} \\
    \cmidrule(lr){4-7} \cmidrule(lr){8-10}
    Cell & Model & Prev. & Train & Val & Val@test & Test & Train & Val & Test \\
    \midrule
    Panel $q_1$ & GraphSAGE      & 34/45/28 & .653 & .815 & .694 & .522 & .746 & .853 & .673 \\
                & XGBoost        &                            & .776 & .829 & .717 & .464 & .846 & .845 & .677 \\
    Panel $q_2$ & GAT            & 34/45/28 & .718 & .843 & .739 & .513 & .799 & .866 & .679 \\
                & XGBoost        &                            & .877 & .885 & .817 & .452 & .915 & .876 & .650 \\
    Panel $q_3$ & GCN            & 34/45/28 & .785 & .860 & .769 & .460 & .863 & .863 & .679 \\
                & FT-Transformer &                            & .880 & .903 & .837 & .508 & .919 & .906 & .686 \\
    Full $q_1$  & GAT            & 43/56/33 & .794 & .906 & .803 & .626 & .810 & .888 & .721 \\
                & Logistic reg.  &                            & .672 & .894 & .776 & .569 & .717 & .884 & .728 \\
    Full $q_2$  & GraphSAGE      & 38/53/30 & .699 & .883 & .769 & .596 & .764 & .886 & .725 \\
                & FT-Transformer &                            & .700 & .895 & .794 & .476 & .756 & .879 & .674 \\
    Full $q_3$  & GCN            & 30/45/27 & .778 & .828 & .707 & .493 & .866 & .874 & .693 \\
                & MLP            &                            & .853 & .898 & .827 & .495 & .908 & .899 & .688 \\
    \bottomrule
  \end{tabular}
\end{table*}

\end{document}